# parallelMCMCcombine: An R Package for Bayesian Methods for Big Data and Analytics

Alexey Miroshnikov[1], Erin M. Conlon[1*]

1 Department of Mathematics and Statistics, University of Massachusetts, Amherst, Massachusetts, United States of America

* E-mail: econlon@mathstat.umass.edu

# Abstract

Recent advances in big data and analytics research have provided a wealth of large data sets that are too big to be analyzed in their entirety, due to restrictions on computer memory or storage size. New Bayesian methods have been developed for large data sets that are only large due to large sample sizes; these methods partition big data sets into subsets, and perform independent Bayesian Markov chain Monte Carlo analyses on the subsets. The methods then combine the independent subset posterior samples to estimate a posterior density given the full data set. These approaches were shown to be effective for Bayesian models including logistic regression models, Gaussian mixture models and hierarchical models. Here, we introduce the R package `parallelMCMCcombine` which carries out four of these techniques for combining independent subset posterior samples. We illustrate each of the methods using a Bayesian logistic regression model for simulation data and a Bayesian Gamma model for real data; we also demonstrate features and capabilities of the R package. The package assumes the user has carried out the Bayesian analysis and has produced the independent subposterior samples outside of the package. The methods are primarily suited to models with unknown parameters of fixed dimension that exist in continuous parameter spaces. We envision this tool will allow researchers to explore the various methods for their specific applications, and will assist future progress in this rapidly developing field.

# Introduction

In this age of big data and analytics, statisticians are facing new challenges due to the exponential growth of information being produced. Here, big data refers to data sets that are too large and complex for classic analysis tools to be used. An extensive number of application areas are affected by big data, including genomics, healthcare, energy, finance, sustainability and meteorology. One primary difficulty in analyzing these large data sets is the restriction on file sizes that can be read into computer memory (RAM); in addition, it may be necessary to store and process data sets on more than one machine due to their large sizes. Several recent Bayesian and Markov chain Monte Carlo (MCMC) methods have been developed to address these issues, where data sets are large only due to large sample sizes. One approach partitions large data sets into smaller subsets, and parallelizes the MCMC computation by analyzing the subsets on separate machines (Langford et al. [1]; Newman et al. [2], Smola et al. [3]); here, information is exchanged at each iteration of the Markov chains, requiring communication between machines. Due to the slow performance of these techniques, alternative methods have been introduced that do not require communication between machines (Neiswanger et al. [4], Scott et al. [5]). These recent methods divide the data into subsets, perform Bayesian MCMC computation on the subsets, and then combine the results back together; the separate analyses are run independently, and are thus parallel, communication-free methods. Specifically, Neiswanger et al. [4] introduces several kernel density estimators that approximate the posterior density for each data subset; the full data posterior is then estimated by multiplying the subset posterior densities together. Alternatively, Scott et al.

[5] developed methods that combine subset posteriors into the approximated full data posterior using weighted averages over the subset MCMC samples.

Previous research in communication-free parallel computing for MCMC methods involved copying the full data set to each machine and computing independent, parallel Markov chains on each machine (Wilkinson [6], Laskey and Myers [7], Murray [8]). However, these methods are not appropriate when the full data set is too large to be read into computer memory. Several additional methods for computation with large data sets in a Bayesian setting have been introduced, but each has limitations. Rue et al. [9] introduced Integrated Nested Laplace Approximation (INLA) for big data, but this method has computational expense that increases exponentially with the number of unknown model parameters. Other approaches include importance resampling methods (Huang and Gelman [10]), but these strategies have drawbacks in that they can collapse to a single point in high dimensional parameter spaces.

The subset-based parallel communication-free MCMC methods hold great promise for the future of Bayesian big data analysis and analytics research. Here, we introduce the R package **parallelMCMCcombine** for implementing four of these methods, including those of Neiswanger et al. [4] and Scott et al. [5]. Note that the package assumes that the user has produced the independent subposterior MCMC samples by carrying out the Bayesian analysis outside of the R [11] package, either within R or in a separate software package such as WinBUGS [12], JAGS [13], or Stan [14,15]; the user then reads the results into our R package. The methods are best suited to models with unknown parameters of fixed dimension in continuous parameter spaces (Neiswanger et al. [4], Scott et al. [5]). The `parallelMCMCcombine` package is implemented in R (R Development Core Team 2014 [11]) and is available from the Comprehensive R Archive Network at http://CRAN.R-project.org/. Our paper is organized as follows. In the Methods section, we introduce each of the four methods. In the following section, we describe the main functions and features of the package; we also demonstrate the package using both simulated and real data sets. We summarize our work in the Discussion section.

## Methods

For Bayesian models, the posterior distribution for unknown model parameters $\theta$ given the full data set is the following: $p(\theta | y) \propto p(y | \theta) p(\theta)$. Here, $\theta$ is a $d$-dimensional vector, where $d$ is the number of unknown model parameters, $p(y | \theta)$ is the likelihood of the full data set given $\theta$, and $p(\theta)$ is the prior distribution of $\theta$. Here, $y$ is a set of $n$ data points that are conditionally independent given $\theta$; we assume $y$ is too large to analyze directly, and we thus partition $y$ into non-overlapping subsets $y_m, m = 1, ..., M$. Here, the partition of $y$ is by the samples $n$, such that if $y$ is dimension $n \times \psi$, then the partition is by the following:

$$y = \begin{pmatrix} y_1 \\ y_2 \\ \vdots \\ y_m \end{pmatrix}, \qquad (1)$$

where each $y_m, m=1,...,M$ has $\psi$ columns. In order to estimate $p(\theta|y)$, we sample from each posterior distribution of $\theta$ given the data subset $y_m, m=1,...,M$; these samples are labeled subposterior samples. The subposterior samples are then combined to approximate the full data posterior distribution. Specifically, the steps are as follow:

1) Partition the data $y$ into disjoint subsets $y_m, m=1,...,M$ as described earlier.

2) For $m=1,...,M$, sample from the subposterior density $p_m$, where

$$p_m(\theta) = p_m(\theta|y_m) \propto p(y_m|\theta) p(\theta)^{1/M} \qquad (2)$$

3) The samples from the subposterior densities are combined, assuming independence, to produce samples from an estimate of the subposterior density product $p_1 p_2 \cdots p_M$; this is proportional to the posterior distribution using the full data set, as follows:

$$p(\theta|y) \propto p_1 p_2 \cdots p_M(\theta) \propto \prod_{m=1}^{M} p(y_m|\theta) p(\theta)^{1/M} \qquad (3)$$

Note that the subsets of data $y_m$ are assumed to be conditionally independent across subsets, given the unknown model parameters. Note also that the prior distribution $p(\theta) = \prod_{m=1}^{M} p(\theta)^{1/M}$, so that the total amount of prior information is equivalent in the independent product model and the full-data model. In the following sections, we assume that subposterior draws of $\theta_t^m = (\theta_{t1}^m, \theta_{t2}^m, ..., \theta_{td}^m)$, for machine $m$, $m=1,...,M$, and MCMC iteration $t$, $t=1,...,T$, have been sampled from each of the subposterior densities $p_m(\theta) \propto p(y_m|\theta) p(\theta)^{1/M}$, $m=1,...,M$, outside of our R package. We describe each method for combining the independent subposterior samples across the machines, as well as the R package implementation of each of the methods.

**Average of subposterior samples method**

For subposterior sample $\theta_t^m$, for machine $m$, $m = 1,...,M$, and MCMC iteration $t$, $t = 1,...,T$, we combine the independent subposterior samples into the pooled posterior samples $\theta_t$, $t = 1,...,T$, by averaging the subposterior samples across machines within each iteration $t$, as follows:

$$\theta_t = \frac{1}{M} \sum_{m=1}^{M} \theta_t^m . \tag{4}$$

Here, the covariance between individual model parameters is assumed to be zero.

**Consensus Monte Carlo method, assuming independence of individual model parameters**

Consensus Monte Carlo methods were developed by Scott et al. [5]. The first Consensus Monte Carlo method assumes the unknown model parameters are independent, and combines the independent subposterior samples across machines within each iteration into the pooled posterior samples $\theta_t = (\theta_{t1}, \theta_{t2}, ..., \theta_{td})$, $t = 1,...,T$, using a weighted average, as follows:

$$\theta_{ti} = \frac{\left(\sum_{m=1}^{M} W_{mi} \theta_{ti}^m\right)}{\left(\sum_{m=1}^{M} W_{mi}\right)}, \quad i = 1,...,d . \tag{5}$$

Here, $W_{mi}$ are the weights defined by $W_{mi} = \left(\Sigma_i^m\right)^{-1}$, where the quantity $\Sigma_i^m = Var(\theta_i | \mathbf{y}_m)$ is estimated by the sample variance of the $T$ MCMC samples $\theta_{1i}^m, \theta_{2i}^m, ..., \theta_{Ti}^m$ of the *i*-th component on the *m*-th machine. Scott et al. [5] state that for models with a large number of unknowns, the user may prefer to ignore the covariances between the model parameters and use this independence method instead of the method that takes into account the covariances between model parameters, which is described next.

**Consensus Monte Carlo method, assuming covariance among model parameters**

The second Consensus Monte Carlo method introduced by Scott et al. [5] assumes that the model parameters are correlated. This method combines the independent subposterior samples across machines within an iteration into the pooled posterior samples $\theta_t, t = 1,...,T,$ using weighted averages, as follows:

$$\theta_t = \left(\sum_{m=1}^M W_m\right)^{-1} \left(\sum_{m=1}^M W_m \theta_t^m\right). \qquad (6)$$

Here, $W_m = \Sigma_m^{-1}$ for each machine *m*, $m = 1,...,M$, where $\Sigma_m = Var(\theta | \mathbf{y}_m)$ is the variance-covariance matrix for the *d* unknown model parameters and has dimension $d \times d$; $\Sigma_m$ is estimated by the sample variance-covariance matrix of the *T* MCMC subposterior samples: $\theta_1^m, \theta_2^m, ..., \theta_T^m$.

Examples of data sets for the consensus Monte Carlo method that assumes covariance among model parameters would be multiple regression models where the predictors are correlated with each other. In this type of data set, the model parameters are the multiple regression coefficients, which we do not assume to be independent. The consensus Monte Carlo methods assume that the subsets of data $\mathbf{y}_m$ are conditionally independent across subsets, given parameters, but it is possible to have dependence structure between the elements within each subset $y_m$. For this, Scott et al. [5] gives an example of a model with data that has a nested structure, and where the consensus Monte Carlo method can be applied to the hyperparameters; this is given in Section 3.4.1 entitled "Nested hierarchical models". For this example, the data needs to be partitioned so that no specific group is split across subsets $y_m$. An example of this type of data is in statistical genomics, with analysis of gene expression data from different laboratories, where the laboratories represent groups. The data from a specific laboratory needs to be kept within a single data subset $y_m$ and not split across multiple subsets. Another example would be medical clinics, where a health outcome for patients is being measured. The model would keep the data for an individual clinic within a single data subset $y_m$ and not split the data across multiple subsets.

Note that Consensus Monte Carlo methods are exact only for Gaussian posterior distributions. However, the methods are also applicable for non-Gaussian posterior distributions, based on the Bayesian central limit theorem (Bernstein von-Mises theorem; see Van der Vaart [16], Le Cam and Yang [17]). This theorem states that posterior distributions tend toward Gaussian limits for large samples under standard regularity conditions in asymptotics. Scott et al. [5] demonstrated that their methods work well for specific Bayesian models with Gaussian posteriors as well as some Bayesian models with non-Gaussian posteriors, for both simulation and real data sets. See the Discussion section for further applicability of the consensus Monte Carlo methods.

**Semiparametric density product estimator method**

The next method involves kernel density estimators, and was developed by Neiswanger et al. [4]. Here, subposterior densities for each data subset are estimated using kernel smoothing techniques; the subposterior densities are then multiplied together to approximate the posterior density based on the full data set. For the semiparametric density product estimator method, the subposterior density $p_m(\theta)$ is viewed as a product $p_m(\theta) = r_m(\theta) \hat{f}_m(\theta)$. For this, $\theta$ is again a $d$-dimensional vector, where $d$ is the number of unknown model parameters. Here, $\hat{f}_m(\theta)$ is a parametric estimator, and $r_m(\theta) = p_m(\theta)/\hat{f}_m(\theta)$ is a correction function whose nonparametric estimator $\hat{r}_m(\theta)$ is defined as follows:

$$\hat{r}_m(\theta) = \frac{1}{T}\sum_{i=1}^{T} \frac{1}{h^d} K\left(\frac{\theta - \theta_t^m}{h}\right) \frac{1}{\hat{f}(\theta_t^m)}. \tag{7}$$

For this, $K(\cdot)$ is the kernel function, $d$ is the dimension of $\theta$, i.e. the number of unknown model parameters, and $h > 0$ is the bandwidth, which is a smoothing parameter (for details see Hjort and Glad [18] and Neiswanger et al. [4]).

For the semiparametric density estimator of Neiswanger et al. [4], the estimated subposterior density $\hat{p}_m(\theta)$ is defined as the product of the corrected nonparametric estimator $\hat{r}_m(\theta)$ and the parametric estimator $\hat{f}_m(\theta)$. For this, $\hat{r}_m(\theta)$ is defined as above, using the Gaussian kernel $K(\cdot)$ with bandwidth $h > 0$, and $\hat{f}_m(\theta) = N_d\left(\theta \mid \hat{\mu}_m, \hat{\Sigma}_m\right)$, where $\hat{\mu}_m$ and $\hat{\Sigma}_m$ are the sample means and sample covariances, respectively, of the subposterior samples $\theta_1^m, \theta_2^m, ..., \theta_T^m$. The values of $\hat{\mu}_m$ and $\hat{\Sigma}_m$ are defined below in Equation (17). The formula for the estimator is then

$$\hat{p}_m(\theta) = \hat{r}_m(\theta)\hat{f}_m(\theta) \tag{8}$$

$$= \left(\frac{1}{T}\sum_{t=1}^{T}\frac{1}{h^d}\frac{N_d\left(\theta | \theta_t^m, h^2 I_d\right)}{N_d\left(\theta_t^m | \hat{\mu}_m, \hat{\Sigma}_m\right)}\right)\left(N_d\left(\theta | \hat{\mu}_m, \hat{\Sigma}_m\right)\right). \tag{9}$$

Here, in Equation (9), we have substituted the expression $N_d\left(\theta | \theta_t^m, h^2 I_d\right)$ for $K\left(\frac{\theta - \theta_t^m}{h}\right)$ when the Gaussian kernel is used for $K(\cdot)$ from Equation (7). In Equation (9), we have also substituted the expression $N_d\left(\theta_t^m | \hat{\mu}_m, \hat{\Sigma}_m\right)$ for $\hat{f}(\theta_t^m)$ from Equation (7). The $M$ subposteriors are then multiplied together, assuming independence, to form the semiparametric density product estimator:

$$\widehat{p_1 p_2 \cdots p_M}(\theta) = \hat{p}_1 \hat{p}_2 \cdots \hat{p}_M(\theta) \tag{10}$$

$$\propto \sum_{t_1=1}^{T}\sum_{t_2=1}^{T}\cdots\sum_{t_M=1}^{T} W_{t\bullet} N_d\left(\theta | \mu_{t\bullet}, \Sigma_{t\bullet}\right), \tag{11}$$

where $t\bullet = (t_1, t_2, ..., t_M)$ is a vector of indices and

$$W_{t\bullet} = \frac{w_{t\bullet} N_d\left(\bar{\theta}_{t\bullet} | \hat{\mu}_M, \hat{\Sigma}_M + \frac{h^2}{M}I_d\right)}{\prod_{m=1}^{M} N_d\left(\theta_t^m | \hat{\mu}_m, \hat{\Sigma}_m\right)}, \tag{12}$$

$$\bar{\theta}_{t\bullet} = \frac{1}{M}\sum_{m=1}^{M}\theta_{t_m}^m, \tag{13}$$

$$w_{t\bullet} = \prod_{m=1}^{M} N_d\left(\theta_t^m \mid \bar{\theta}_{t\bullet}, h^2 I_d\right), \tag{14}$$

$$\Sigma_{t\bullet} = \left(\frac{M}{h^2} I_d + \hat{\Sigma}_M^{-1}\right)^{-1}, \tag{15}$$

$$\mu_{t\bullet} = \Sigma_{t\bullet} \left(\frac{M}{h^2} I_d \bar{\theta}_{t\bullet} + \hat{\Sigma}_M^{-1} \hat{\mu}_M\right), \tag{16}$$

$$\text{and } \hat{\mu}_M = \hat{\Sigma}_M \left(\sum_{m=1}^{M} \hat{\Sigma}_m^{-1} \hat{\mu}_m\right), \hat{\Sigma}_M = \left(\sum_{m=1}^{M} \hat{\Sigma}_m^{-1}\right)^{-1}. \tag{17}$$

The semiparametric density product estimator can be viewed as a mixture of $T^M$ Gaussian densities, with mixture weights $W_{t\bullet}$. Neiswanger et al. [4] outline an algorithm for generating samples from probability density function (10) by first sampling a particular mixture component and then sampling from the selected mixture component. The sampling is carried out using an independent Metropolis within Gibbs (IMG) sampler, which is a type of MCMC sampler. For the IMG sampler, at each iteration, a new mixture component is proposed by uniformly sampling one of the M indices $t_m \in t\bullet = (t_1, t_2, ..., t_M)$. This proposed component is then either accepted or rejected based on its mixture weight. This method uses a procedure that changes the bandwidth $h$ for each iteration $t$; specifically, $h = t^{(-1/(4+d))}$, $t = 1, ..., T$, so that $h$ decreases at each iteration $t$ in the algorithm. This procedure is referred to as *annealing* by Neiswanger et al. [4]. Annealing is used for the bandwidth $h$ since it typically allows for a more extensive search of the sample space compared to fixed values of the bandwidth. With annealing, by design, the algorithm begins with a large probability of accepting many low-probability solutions; this probability of acceptance decreases as $h$ decreases, resulting in fewer low-probability solutions being accepted as h decreases. The algorithm for the semiparametric density product estimator creates the combined posterior samples $\theta_t$, $t = 1,...,T$.

Neiswanger et al. [4] demonstrated the semiparametric density product estimator for several models with simulation data, including a Bayesian logistic regression model, a Bayesian Gaussian mixture model and a Bayesian hierarchical Poisson-Gamma model. The methods are also implemented using real-world data for a Bayesian logistic regression model. The theoretical results for the semiparametric density product estimator are also applicable to Bayesian generalized linear models such as Bayesian linear regression and Bayesian Poisson regression, and Bayesian finite-dimensional graphical models with no constraints on the parameters. See the Discussion section below for further applicability of the semiparametric density product estimator method.

**A metric for comparing densities**

Results for the four methods described above can be compared in $\mathbb{R}$ using an estimate of the $L_2$ distance, $d_2(p, \hat{p})$, between the full data posterior $p$ and the combined estimated posterior $\hat{p}$ (as introduced in Neiswanger et al. [4]), where

$$d_2(p, \hat{p}) = \|p - \hat{p}\|_{L_2} = \left( \int_{\mathbb{R}^d} \left( p(\theta) - \hat{p}(\theta) \right)^2 d\theta \right)^{1/2}. \tag{18}$$

We use MCMC samples and kernel density estimation for each posterior, as described in Neiswanger et al. [4] and Oliva et al. [19]. The estimated relative $L_2$ distance, relative to the full data posterior, is reported for each of the four methods in the data examples in the following section.

## Using Package parallelMCMCcombine

### Package overview

The R package **parallelMCMCcombine** assumes the user has run the Bayesian models for the subset data either within R [11] or in a separate software package such as WinBUGS [12], JAGS [13], or Stan [14,15]. The user then reads the MCMC results into an array in R, with dimension specified below. The **parallelMCMCcombine** package has four major functions, based on the four methods described above in the Methods section, respectively: **sampleAvg()**, **consensusMCindep()**, **consensusMCcov()** and **semiparamDPE()**. The arguments used in a to call to the function **sampleAvg()** are the following; the same arguments are used for **consensusMCindep()**, **consensusMCcov()**:

```
sampleAvg(subchain=NULL, shuff=FALSE)
```

The descriptions of the available arguments are:

**subchain**  An array with dimensions = $(d, T, M)$, where $d$ = number of unknown model parameters, $T$ = number of MCMC samples, $M$ = number of subsets of the full data. This is the input data that is input by the user; these are the Bayesian MCMC subposterior samples produced outside of our R package.

**shuff** a logical value indicating whether the *d*-dimensional samples within a machine should be randomly permuted. The purpose of this option is to remove potential correlations between MCMC samples from different machines.

The arguments used in a to call to the function **semiparamDPE()** are the following:

```
semiparamDPE(subchain=NULL, bandw=rep(1.0,
 dim(subchain)[1]), anneal=TRUE, shuff=FALSE)
```

The **subchain** and **shuff** arguments are the same as above; the descriptions of the remaining arguments are:

**bandw** the vector of bandwidths *h* of length *d* to be specified by the user, where *d* = number of unknown model parameters. Here, bandwidths are the tuning parameters used in kernel density approximation employed by the semiparametric density product estimator method. The default value is a vector of 1's of length *d* (see Appendix for more detail).

**anneal** a logical value. If **TRUE**, the bandwidth **bandw** (instead of being fixed) is decreased for each iteration of the algorithm (referred to as *annealing*) as $h_i$ = **bandw** $* t^{(-1/(4+d))}, i = 1,...,d; t = 1,...,T$; $d$ = number of unknown model parameters, as described above in the Methods section and in Algorithm 1 of Neiswanger et al. [4]. If **FALSE**, the bandwidth vector *h* is fixed as *h* = **bandw** (see Appendix for more detail).

Note that the default values for the **bandw** and **anneal** arguments are equivalent to the algorithm of Neiswanger et al. [4] described above in the Methods section, which again specifies that $h = t^{(-1/(4+d))}$, $t = 1,...,T$ (see Appendix for more details). The user is given the option to change both the **bandw** and **anneal** arguments, so that the bandwidth values $h_i$, $i = 1,...,d$ can be either fixed at a different value or can decrease with each iteration $t$ (i.e. annealed) with a different starting value. See the Appendix for further detail on kernels and bandwidth selection.

The returned value of each of the four functions described above is a matrix with dimension = $(d,T)$, where $d$ = number of unknown model parameters, $T$ = number of input MCMC samples. The values within the matrix are the combined posterior samples based on the selected function.

**Example: Bayesian logistic regression model for simulation data**

Logistic regression is an extensively used method in many application areas for the analysis of categorical data, and is also used for classification purposes. Here, we generate simulation data for logistic regression and carry out a Bayesian logistic regression analysis in order to demonstrate the implementation of our R package.

*Simulation data for logistic regression*

We simulated 100,000 observations from a logistic regression model with five covariates; the sample size of 100,000 was chosen so that a full data analysis was still feasible. The covariates $X$ and model parameters $\beta$ were simulated from standard normal distributions.

The resulting simulated values of $\beta$ were: $\beta = (0.47, -1.70, 0.54, -0.90, 0.86)'$, which are the parameters estimated in our analysis. The outcome values $y_i$, $i = 1,...,100,000$, were then simulated from the following:

$$y_i \sim \text{Bernoulli}(p_i), \tag{19}$$

$$p_i = \frac{\exp(X_i\beta)}{1+\exp(X_i\beta)}, \tag{20}$$

where $X_i$ denotes the $i$th row of $X$ (see also Neiswanger et al. [4] and Scott et al. [5]).

Our Bayesian logistic regression model with five covariates is the following:

$$y_i \mid p_i \sim \text{Bernoulli}(p_i), \ i = 1,...,n, \tag{21}$$

$$\text{logit}(p_i) = \beta_1 x_{i1} + \beta_2 x_{i2} + \beta_3 x_{i3} + \beta_4 x_{i4} + \beta_5 x_{i5}, \tag{22}$$

where

$$\text{logit}(p_i) = \log\left(\frac{p_i}{1-p_i}\right). \tag{23}$$

We assigned uninformative priors to the $\beta$ parameters (see Carlin and Louis [20], Gelman et al. [21], Liu [22]) as follows:

$$p(\beta_i) \propto 1, \ i = 1,...,5. \tag{24}$$

The full data set was divided into 10 subsets of size 10,000 each. The WinBUGS [12] software package was implemented for MCMC sampling of the $\beta_i$ model parameters for each of the data subsets; we sampled 50,000 iterations after burnin of 2,000 iterations for each. We also implemented the Bayesian model for the full data set for the $\beta_i$ model parameters for comparison, again for 50,000 iterations each after burnin of 2,000 iterations. Results are described in the following sections. Note that for the subset analyses and the full data analysis, the same uninformative prior distributions are used for the $\beta_i$ model parameters. No prior adjustment is necessary, taking into account the number of subsets, since the subset and full data sample sizes are large and the prior distributions are uninformative (see Scott et al. [5]).

*Illustration of R package for Bayesian logistic regression*

For the logistic regression model of the previous section, the MCMC output from the separate machine analyses is read into an array named **logistic.MCMC** in R and has array dimension = (5,50000,10), for 5 unknown model parameters, 50,000 MCMC samples and 10 data subsets. At the R prompt, the user enters the following command for the method of averaging subposterior samples; here, we permute the *d*-dimensional samples within each machine:

```
> logistic.sampleavg.combine <- sampleAvg(logistic.MCMC,
shuff=TRUE)
```

The output **logistic.sampleavg.combine** is a matrix with dimensions = (5,50000) for 5 unknown model parameters and 50,000 combined posterior samples $\theta_t$, $t = 1,...,T$ (= input number of samples). We plot the estimated combined posterior density and full data posterior density in Figure 1 for the marginal of the first unknown model parameter $\beta_1$; we also plot the 10 marginal subposterior densities. In this figure, the combined estimated posterior density is similar to the full data posterior density, with estimated relative $L_2$ distance of 0.034 (see Table 1). Similar results were found for the marginal densities of the remaining four parameters $\beta_2,...,\beta_5$ (not shown). Note that the MCMC output file for the full data posterior density is not used within the R package.

The procedure for carrying out the two combining methods **consensusMCindep()** and **consensusMCcov()** is similar to the procedure shown above for **sampleAvg()**. For each of these methods, we show results similar to those described above in Figure 1. These methods showed improvement versus the sample average method, with estimated relative $L_2$ distance values of 0.024 for the consensus Monte Carlo independence method, and 0.015 for the consensus Monte Carlo covariance method (see Table 1). Similar results were found for the marginals for $\beta_2,...,\beta_5$ (not shown).

For the semiparametric density product estimator method, the command is similar to that shown above, but with additional arguments. Here, we use the default values for the arguments **bandw** and **anneal** that are equivalent to the algorithm in Neiswanger et al. [4] described above; we also permute the *d*-dimensional samples within each machine:

```
> logistic.semiparamDPE.combine <-
semiparamDPE(logistic.MCMC, bandw=rep(1.0,
dim(subchain)[1]), anneal=TRUE, shuff=TRUE)
```

The output `logistic.semiparamDPE.combine` is a matrix with dimensions = (5,50000) for 5 unknown model parameters and 50,000 combined posterior samples $\theta_t$, $t = 1,...,T$ (= input number of samples). We again plot the estimated combined posterior density and the full data posterior density in Figure 1 for the marginal for $\beta_1$. The estimated relative $L_2$ distance of 0.046 is larger than the previous three methods (see Table 1). We note that the semiparametric density product estimator method is sensitive to the choice of bandwidth. We found that changing the `anneal` argument from `TRUE` to `FALSE`, leaving all other previous arguments the same, improved the estimated relative $L_2$ distance for this model and data set. This option of `anneal=FALSE` is equivalent to setting the bandwidth $h = 1$ in the algorithm of Neiswanger et al. [4]. Figure 1 shows improved results, with the estimated relative $L_2$ distance decreasing from 0.046 to 0.020 (see also Table 1). We found similar results for the marginal densities of the remaining four parameters $\beta_2,...,\beta_5$ (not shown). In summary, of the four methods, for this model and simulation data set, the consensus Monte Carlo covariance method produced the smallest estimated relative $L_2$ distance, followed by the semiparametric density product estimator method, the consensus Monte Carlo independence method and the sample average method.

**Example: Bayesian Gamma model for real data**

Here, we analyze real data for all commercial flights within the United States for the three-month period November 2013 through January 2014, obtained from the U.S. Department of Transportation [23]. The random variable of interest is the arrival delay in minutes for each flight; this is defined as the difference between the scheduled and actual arrival time. Values of fifteen minutes and lower are considered on-time arrivals (following guidelines of [23]), with 329,905 remaining data values for the arrival delays. We applied a square root transformation to the data set; the resulting transformed data values followed an approximate Gamma distribution. Our model for the transformed data values $y_i$ is then as follows:

$$y_i \mid \alpha, \beta \sim \text{Gamma}(\alpha, \beta), \quad i = 1, \ldots, n. \tag{25}$$

We estimate the $\alpha$ and $\beta$ parameters using a Bayesian Gamma model; for this, we reparameterize the Gamma distribution in terms of the mean and variance in order to remove correlation between the $\alpha$ and $\beta$ parameters (see Kruschke [24]). Here, we use the following:

$$\alpha = \frac{\lambda^2}{\delta^2}, \tag{26}$$

$$\beta = \frac{\lambda}{\delta^2}, \tag{27}$$

where

$$\lambda = \frac{\alpha}{\beta} = \text{mean of Gamma distribution}, \tag{28}$$

$$\delta^2 = \frac{\alpha}{\beta^2} = \text{variance of Gamma distribution}. \tag{29}$$

We assigned uninformative prior distributions to $\lambda$ and $\delta$, as follows:

$$\lambda \sim \text{Uniform}(0.0001, 10000), \tag{30}$$

$$\delta \sim \text{Uniform}(0.0001, 10000). \tag{31}$$

The full data set was divided into five subsets of size 65,981 each. Using WinBUGS, we sampled the posterior distributions of the $\alpha$ and $\beta$ parameters for each of the data subsets as well as the full data set, for 50,000 iterations after burnin of 2,000 iterations. Results are described in the following sections. Note again that the same uninformative prior distributions are used for the subset analyses and the full data analysis. Prior adjustment is not necessary, accounting for the number of subsets, since the subset and full data sample sizes are large and the prior distributions are uninformative (see Scott et al. [5]).

*Results of R package for real data*

The WinBUGS output for the Bayesian Gamma model for the airlines data is read into R as an array named "airlines.MCMC" with dimension = (2,50000,5), for 2 model parameters, 50,000 MCMC samples and 5 data subsets. We repeat the commands shown for the R package for the logistic regression model of the previous sections, including analyzing the two possible values `TRUE` and `FALSE` of the `anneal` argument for the semiparametric density product estimator function `semiparamDPE()`. Figure 2 shows results for each of the four methods for the parameter $\alpha$; plots are also displayed for the full data posterior density and five subposterior densities. For this model and data set, all four methods create combined estimated posterior densities that are similar to the full data posterior density, with estimated relative $L_2$ distances ranging from 0.016 to 0.024 (Table 1). For the semiparametric density product estimator method, we again found improved performance when setting `anneal=FALSE` so that the bandwidth value is fixed with $h = 1$ in the algorithm of Neiswanger et al. [4]; however, the improvement was not large (see Figure 1 and Table 1). To summarize, the consensus Monte Carlo covariance method generated the smallest estimated relative $L_2$ distance, followed by the consensus Monte Carlo independence method, the semiparametric density product estimator method and the sample average method. Similar results were found for the $\beta$ model parameter (not shown).

**Software computational time**

Here, we show the computational time for the four methods, using the number of model parameters $d$ = 2, 5, 10 and 50, and the number of subsets $M$ = 5, 10, 20 and 100, and number of MCMC samples = 50,000 (see Table 2). For this, we created simulation MCMC output, which is described below in "Simulation procedure". We also show computational times for our data examples, for the combinations $d$ = 2, $M$ = 5 (Gamma model) and $d$ = 5, $M$ = 10 (logistic regression model). The semiparametric density product estimator method has the longest computational time, which is up to several hundred times longer than the consensus Monte Carlo covariance method and up to several thousand times longer than the fastest method, which is the sample average method. Note that the computational time results for the simulation data are within seconds of the computational times for the two example data sets for the same values of $d$ and $M$ (Table 2).

Users may also want to consider the computational time for producing the MCMC samples, which we assume the user produces outside of our R package using either R or Bayesian software such as WinBUGS [12], JAGS [13] or Stan [14,15]. The MCMC samples are then used as input to the **parallelMCMCcombine** package. In Table 3, we include the computational times for producing the 52,000 MCMC samples (including burnin) for the full data sets and the data subsets for our two data examples, which we produced using WinBUGS [12]. We also show additional computational times for different numbers of subsets of $M = 5$, 10 and 20. Note that the computational time for individual subsets decreases as $M$ increases, since the number of samples is smaller in each subset, but there are more subsets to run as $M$ increases. The total computational time for the subset data versus the full data set are comparable; in our two examples, the full data set had longer total computational times than the subset data, but the difference is not always large (see Table 3).

*Simulation procedure*

We simulated the MCMC samples to determine computational time as follows. For each combination of number of model parameters $d$ and number of subsets $M$, we randomly generated a mean for each unknown model parameter $i$, $i = 1,…,d$ by sampling uniformly in the range [-200,200]. We then generated the 50,000 samples for each of the subset posteriors $m$, $m = 1,…,M$ for each unknown model parameter $i$ by sampling from a normal distribution with this sampled mean and variance $= 2$.

## Discussion

Here, we introduce and demonstrate the R package **parallelMCMCcombine** for Bayesian analysis of large-scale data sets, which are only large due to large sample sizes. The package includes four methods for combining independent subset posterior samples into combined estimated posterior densities for unknown Bayesian model parameters. The independent subset posterior samples for each of the four methods are assumed to be generated by parallel, communication-free MCMC sampling techniques applied to the data subsets; these subset posterior samples are assumed to be produced outside of our R package. For further analysis beyond our R package, users can compare the four methods using various metrics. We showed comparisons using estimated relative $L_2$ distances between the combined estimated posterior densities and the full data posterior density, when a full data analysis is possible. We found that the four methods performed differently in terms of which produced the smallest $L_2$ distance, depending on the models and data sets. Further research is needed to determine conditions under which each method performs optimally, including the number of data subsets, the types of models and the number of unknown model parameters, and the number of MCMC samples, among other variables. In addition, the semiparametric density product estimator method is sensitive to the choice of bandwidth, and more study is needed in the area of bandwidth selection. The `parallelMCMCcombine` package is designed to help investigators explore the various methods for their specific applications and to assist in the development of further research for this rapidly expanding field.

The four methods included in our R package are best suited to models with unknown parameters with fixed dimension in continuous parameter spaces (Neiswanger et al. [4], Scott et al. [5]). Models with label switching, changing dimension numbers, and model averaging with spike and slab priors are not well suited to the methods included in our R package. The methods also do not work well for discrete parameters. There are open questions regarding the applicability of the methods that need further exploration, as discussed in Scott et al. [5]; these include hierarchical models with crossed random effects and Gaussian processes that have covariance functions that are not trivial.

For the number of unknown model parameters *d* that is appropriate for the above four methods, Neiswanger et al. [4] illustrates the semiparametric density product estimator method using a synthetic data set with $d = 50$, and a real-world data set with $d = 54$ for logistic regression. They performed a simulation study for logistic regression to show that the performance of the semiparametric density product estimator method scales well with dimension, with a maximum $d = 130$. They also use a Gaussian mixture model where the data was sampled from a ten component mixture of 2-dimensional Gaussians. Scott et al. [5] illustrate their method using a multivariate normal model with unknown mean and variance; here, the unknown mean is dimension 5×1, and the unknown variance is dimension 5×5. They also illustrate their method using a logistic regression model with $d = 5$. Similar to the recommendations of Scott et al. [5], we recommend that the user work with simulation data for their particular applications to determine conditions under which the four methods perform well. Note that there is virtually no limitation on the dimension *d* to be used in the R package, except for the limits of the computer memory of the user.

The semiparametric density product estimator method has the longest computational time of the four methods of our R package. The direct use of *d*-dimensional multivariate Normal distributions slows down computations considerably for this procedure. We are currently working to optimize the algorithm to improve the speed; this will be implemented in the next release of the package.

# Appendix

**Remarks on kernels and bandwidth selection**

Given a kernel $K(\bullet)$ in $d$ dimensions, the expression $\frac{1}{h^d} K\left(\frac{\theta}{h}\right)$ in formulas for nonparametric and parametric estimators can be replaced by $|H|^{-1/2} K(H^{-1/2}\theta)$, where $H$ is the bandwidth $d \times d$ matrix which is symmetric and positive definite. In this case, for the semiparametric estimator, in the formula for the weights $W_{t\bullet}$, the matrix $h^2 I_d$ will be replaced by $H$. Note that the default choices for the arguments `bandw` and `anneal` in the R package function `semiparamDPE()` are equivalent to setting the bandwidth $h$ as it is defined in the semiparametric density product estimator algorithm of Neiswanger et al. [4], which is $h = t^{(-1/(4+d))}$, $t = 1,...,T$, where $T$ is the number of samples.

When using the semiparametric density product estimator method, often the choice of $H$ is crucial in the estimation. In particular, if the bandwidth is a fixed diagonal matrix $H = \text{diag}(h_1^2, h_2^2, ..., h_d^2)$, the smoothing parameters $h_i$ may be chosen according to Silverman's rule of thumb

$$h_i = \left(\frac{4}{d+2}\right)^{1/(d+4)} T^{(-1/(d+4))} \sigma_i, \tag{32}$$

where $\sigma_i$ is the standard deviation of the $i$-th scalar component of $\theta$, and $T$ is the total number of samples (see Silverman [25], Wand and Jones [26,27] and Duong and Hazelton [28]). This option is included as an example in the `semiparamDPE()` function of the R package.

28. Duong T, Hazelton ML (2003) Plug-in bandwidth matrices for bivariate kernel density estimation. J Nonparametric Statistics 15: 17-30.31

**Figure Legends**

**Figure 1.** Results for the simulation data of the Bayesian logistic regression model, for the marginal of the $\beta_1$ parameter. (a) full data posterior density and 10 subposterior densities for the 10 data subsets; (b)-(f): full data and estimated combined posterior densities for: (b) sample average method; (c) consensus Monte Carlo independence method; (d) consensus Monte Carlo covariance method; (e) semiparametric density product estimator method, with default settings of the function; (f) semiparametric density product estimator method with the same settings as (e) except the argument `anneal=FALSE`. The consensus Monte Carlo covariance method produces the smallest $L_2$ distance (see Table 1).

**Figure 2.** Results for the real airlines data of the Bayesian Gamma model, for the marginal of the $\alpha$ parameter. (a) full data posterior density and 5 subposterior densities for the 5 data subsets; (b)-(f): full data and estimated combined posterior densities for (b) sample average method; (c) consensus Monte Carlo independence method; (d) consensus Monte Carlo covariance method; (e) semiparametric density product estimator method, with default settings of the function; (f) semiparametric density product estimator method with the same settings as (e) except the argument `anneal=FALSE`. The consensus Monte Carlo covariance method produces the smallest $L_2$ distance (see Table 1).

**Tables**

**Table 1.** Estimated relative $L_2$ distances

| Subposterior Samples Combining Method | Bayesian Model | |
|---|---|---|
| | Logistic Regression, $\beta_1$ **parameter** | Gamma, $\alpha$ **parameter** |
| **Sample Average** | 0.034 | 0.024 |
| **Consensus MC Independence** | 0.024 | 0.020 |
| **Consensus MC Covariance** | 0.015 | 0.016 |
| **Semiparametric DPE, `anneal=TRUE`** | 0.046 | 0.022 |
| **Semiparametric DPE, `anneal=FALSE`** | 0.020 | 0.021 |

Estimated relative $L_2$ distances for each of the methods of combining subposterior samples to estimate posterior densities given the full data set. Results are included for the Bayesian logistic regression model with the simulated data set for the marginal densities of the $\beta_1$ parameter, and the Bayesian Gamma model with the real airlines data set for the marginal densities of the $\alpha$ parameter.

**Table 2.** Computational time (in seconds) for the four combining methods

| Number of Subsets M | | Number of Model Parameters d | | | |
|---|---|---|---|---|---|
| | | 2 | 5 | 10 | 50 |
| 5 | Sample Average | 0.06 (0.04) | 0.09 | 0.13 | 0.48 |
| | Consensus Indep | 2 (2) | 2 | 2 | 3 |
| | Consensus Cov | 2 (2) | 2 | 3 | 23 |
| | SemiparamDPE | 401 (402) | 432 | 464 | 1136 |
| 10 | Sample Average | 0.06 | 0.10 (0.12) | 0.20 | 0.89 |
| | Consensus Indep | 2 | 2 (2) | 2 | 4 |
| | Consensus Cov | 4 | 4 (4) | 5 | 36 |
| | SemiparamDPE | 795 | 816 (820) | 880 | 2119 |
| 20 | Sample Average | 0.08 | 0.16 | 0.31 | 1 |
| | Consensus Indep | 2 | 2 | 2 | 7 |
| | Consensus Cov | 6 | 7 | 10 | 70 |
| | SemiparamDPE | 1540 | 1602 | 1729 | 4102 |
| 100 | Sample Average | 0.34 | 0.72 | 1 | 7 |
| | Consensus Indep | 3 | 4 | 5 | 27 |
| | Consensus Cov | 29 | 35 | 47 | 343 |
| | SemiparamDPE | 7522 | 8015 | 8675 | 22540 |

Computational times, in seconds (rounded unless less than 1 second), for the four methods of the R package **parallelMCMCcombine**, using simulation data and $T = 50,000$ MCMC samples. The values in parentheses are for our example data sets; $d = 2$, $M = 5$ is for the Gamma model, and $d = 5$, $M = 10$ is for the logistic model. The results are based on a computer with operating system Windows 7 and an Intel Celeron 1007U CPU 1.5 GHz Processor.

**Table 3.** Computational time (in minutes) for producing the MCMC samples

| Number of subsets $M$ and number of data points | Bayesian Logistic Regression Model, $d = 5$ model parameters | Number of subsets $M$ and number of data points | Bayesian Gamma Model, $d = 2$ model parameters |
|---|---|---|---|
| $M = 5$, 20,000 | 174 per subset, total=5(174)=870 | $M = 5$, 65,981 | 256 per subset, total=5(256)=1280 |
| $M = 10$, 10,000 | 85 per subset, total=10(85)=850 | $M = 10$, 32,991 | 139 per subset, total=10(139)=1390 |
| $M = 20$, 5,000 | 41 per subset, total=20(41)=820 | $M = 20$, 16,496 | 65 per subset, total=20(65)=1300 |
| full data set ($M = 1$), 100,000 | 954 for full data, total=1(954)=954 | full data set ($M = 1$), 329,905 | 1397 for full data, total=1(1397)=1397 |

Average computational times, in minutes (rounded), for producing $T = 52,000$ samples (including burnin) for the data examples, and total computational times. The results are based on the WinBUGS software program and a computer with operating system Windows 7 and an Intel Core i7-4600U CPU 2.1 GHz Processor. Note that the R package **parallelMCMCcombine** is not used to create these samples; the MCMC samples are used as input to the **parallelMCMCcombine** package.

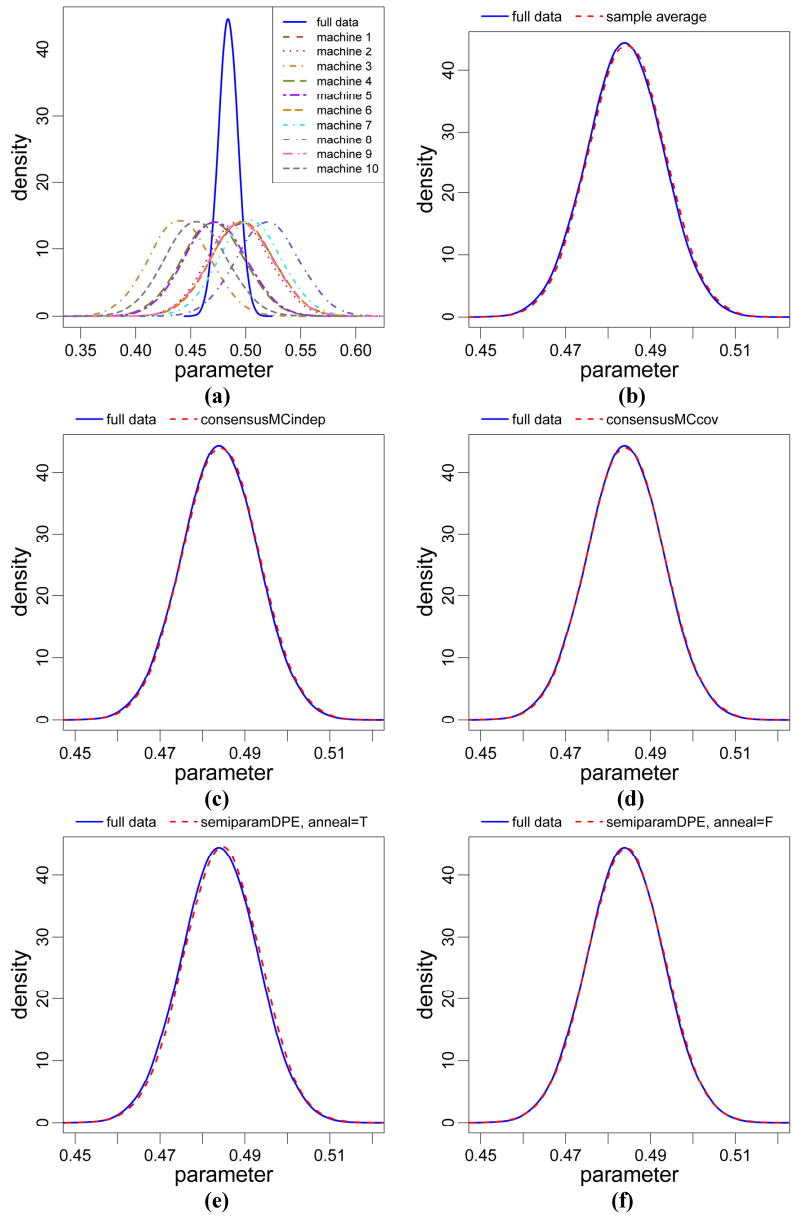

Figure 1

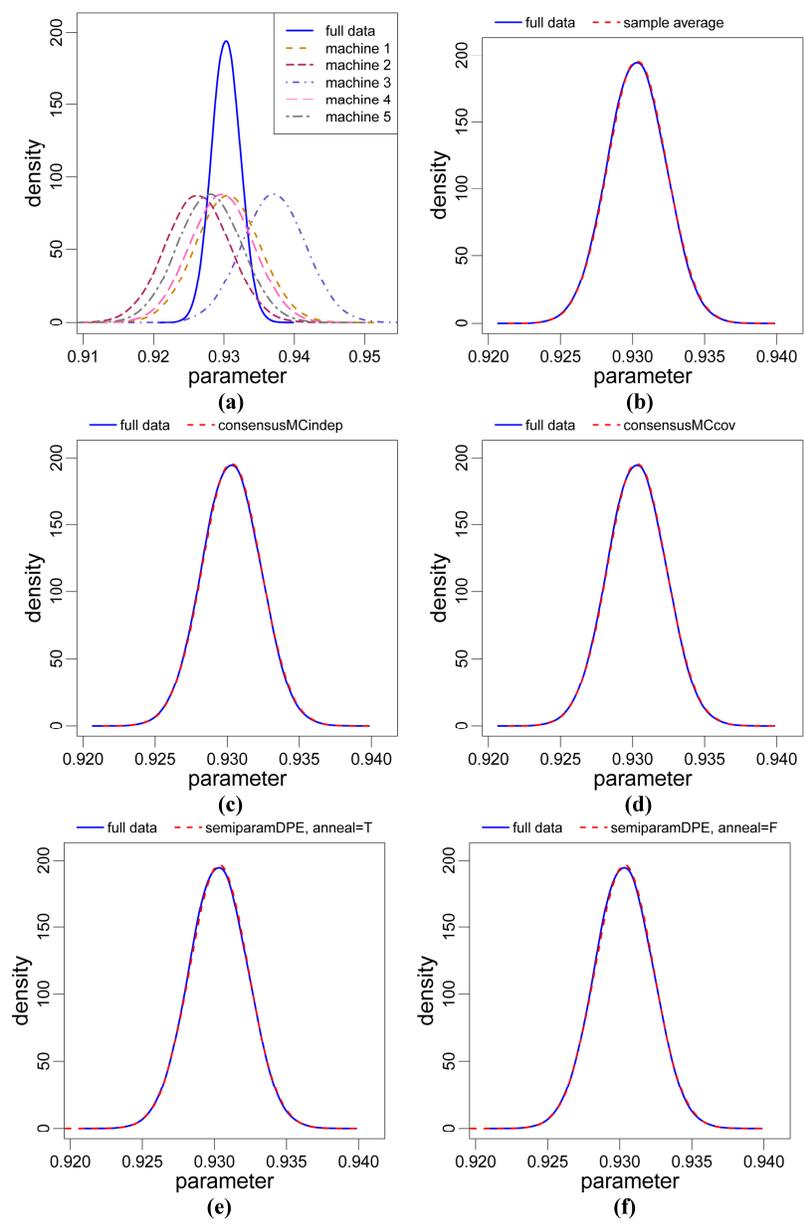

Figure 2